\documentstyle[12pt]{article}
\begin{document}

\title{  Coalescence Model    of 
 Rock-Paper-Scissors  Particles}

\author{Yoshiaki Itoh \thanks{
itoh@ism.ac.jp}\\
Institute of Statistical Mathematics\\ and Graduate University for Advanced Studies\\ Tachikawa, Tokyo  
} 
\maketitle
{\bf Abstract} 
The rock-paper-scissors game, commonly played in East Asia, gives a simple  model to understand 
physical,  biological,  psychological  and other problems. The interacting rock-paper-scissors particle system  is a point of contact between  the kinetic theory of gases by Maxwell and Boltzmann ( collision  model) and the coagulation theory by Smoluchowski ( coalescence model).  A $2s+1$ types extended rock-paper-scissors collision model  naturally introduces a nonlinear integrable system. The time evolution of  the  $2s+1$ types extended rock-paper-scissors coalescence model is obtained from the logarithmic time change of  
the nonlinear integrable  system.  We also discuss the behavior of a discrete rock-paper-scissors coalescence model.

{\bf PACS numbers}: 02.30.Ik,  02.50.Cw, 05.20.Dd

\section{Introduction}
\begin{figure}[h]
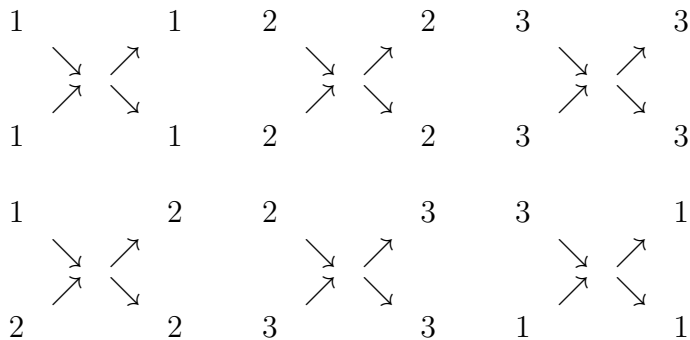

  \[\begin{array}{cccccccccccccccccc}

1&        &        &1&  & &2&        &        &2&  & &3&        &        &3\\
 &\searrow&\nearrow&  & & & &\searrow&\nearrow&  & & & &\searrow&\nearrow& \\
 &\nearrow&\searrow&  & & & &\nearrow&\searrow&  & & & &\nearrow&\searrow& \\
1&        &        &1&  & &2&        &        &2&  & &3&        &        &3\\
 &        &        &  & & & &        &        &  & & & &        &        & \\
1&        &        &2&  & &2&        &        &3&  & &3&        &        &1\\
 &\searrow&\nearrow&  & & & &\searrow&\nearrow&  & & & &\searrow&\nearrow& \\
 &\nearrow&\searrow&  & & & &\nearrow&\searrow&  & & & &\nearrow&\searrow& \\
2&        &        &2&  & &3&        &        &3&  & &1&        &        &1
\end{array} \]
\caption{ Rock-paper-scissors collision model \cite{i71}}\label{collision}
\end{figure}

The rock-paper-scissors game is commonly played in East Asia. The cyclic 
dominance systems  
naturally occurs in biological systems as
 color morphisms of the side-blotched lizard \cite{sl} and strains of Escherichia coli \cite{krfb}.   
The types, rock  paper and scissors,  can also represent social groups, opinions, or survival strategies of organisms in a preferential
attachment graph model \cite{hj} . 
Considering   simple models \cite{ka} for kinetic theory of gases by Maxwell and Boltzmann,  we introduced 
a collision  model   of 
rock (type 1), paper (type 2), and scissors (type 3)  particles with cyclic dominance  Fig. \ref{collision}, where 2 dominates 1, 3 dominates 2, and 1 dominates 3, and    obtained  a Lotka-Volterra equation  \cite{i71} (  the Boltzmann equation for the rock-paper-scissors particles). We can extend the argument to the  $2s+1$ types  rock-paper-scissors particles \cite{i75},    which gives    a nonlinear integrable system \cite{bo,  i87, i09, na} 
 with $s+1$ conserved quantities 
  like    the Toda lattice \cite{to} and  the  Calogero system \cite{ca}. 
 For the case of finite  number of particles,  the probability     of  coexistence of types is  obtained by using $s+1$  martingales, which are  stochastic version of the $s+1$ conserved quantities  \cite{i73, i79}.   
 The rock-paper-scissors lattice model  greatly enriches the dynamics as studied in the physics literatures \cite{fk, fmo, kkwf,  ssi, ta, ti}.

\begin{figure}[h]
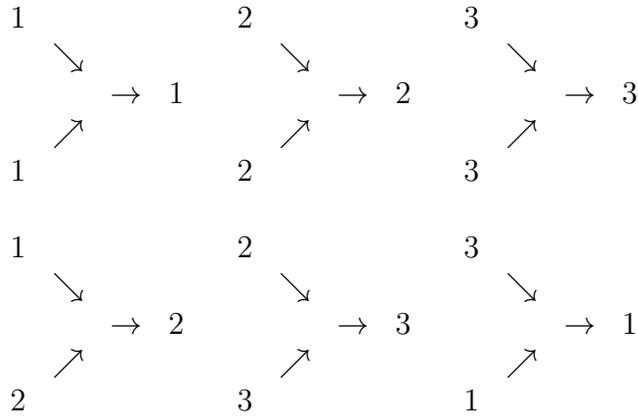

\[\begin{array}{cccccccccccccccccc}

1&        &            & & &2&        &            & & &3&        &           & \\
 &\searrow&            & & & &\searrow&            & & & &\searrow&           & \\
 &        &\rightarrow&1 & & &        &\rightarrow&2 & & &        &\rightarrow&3\\
 &\nearrow&            & & & &\nearrow&            & & & &\nearrow&           & \\
1&        &            & & &2&        &            & & &3&        &           & \\
 &        &            & & & &        &            & & & &        &           & \\          
1&        &            & & &2&        &            & & &3&        &           & \\
 &\searrow&            & & & &\searrow&            & & & &\searrow&           & \\
 &        &\rightarrow&2 & & &        &\rightarrow&3 & & &        &\rightarrow&1\\
 &\nearrow&            & & & &\nearrow&            & & & &\nearrow&           & \\
2&        &            & & &3&        &            & & &1&        &           & \\

\end{array} \]
\caption { Rock-paper-scissors coalescence model \cite{i73}.}\label{coalescence} 
\end{figure}

We introduce a  coalescence model of  rock-paper-scissors particles  as in  Fig.   \ref{coalescence}.  
For a given initial distribution of the particles of rock (type 1), paper  (type 2) and  scissors (type 3), what type of the particle will finally survive?  As we see in Fig. \ref{tree1} and Fig. \ref{tree2} it 
gives  a  leader selection problem, which is for another aspect of  the previously studied  problem\cite{fhi}.    We carried out simulation studies for  the coalescence model of rock-paper-scissors particles  \cite{i73b} 
for finite size fluctuation, where the total number of particles decreases  at each step.  Cyclic   trapping reactions for finite size fluctuations \cite{bk} 
gives  insights to the behavior of our model for finite size. 
 
Here we  study the  time evolution of the coalescence model of rock-paper-scissors particles with sufficiently large number of particles.
 We  apply  the  study of  coalescence of clusters   
\cite{al, krb, sm} to our problem.  Let us follow the argument   \cite{krb} for the simplest case. 
Infinite set of master equations that
describe how the cluster mass distribution $n_k (t)$  evolves by the rule  
 \begin{eqnarray}
A_i+A_j\rightarrow A_{i+j},\label{cluster}
\end{eqnarray} 
 in which clusters of mass $i$ and $j$ irreversibly join  to form a 
cluster of mass $i+j$, is given by 
\begin{eqnarray}  
 \frac{d~n_{k}(t)}{dt}
= 1/2\sum _{i+j=k} n_{i}n_{j}-\sum_{j=1}^{\infty}  n_j n_k. \label{simple}
 \end{eqnarray}
The first term on the right-hand side is  for the creation of a $k$-mers 
due to the coalescence of two clusters of mass $i$ and $j$, we sum over all such pairs with $i+j=k$.  The factor $1/2$ in the gain term is needed to avoid double counting. 

 Instead of Eq. (\ref{cluster}), the rock-paper-scissors coalescence in 
 Fig. \ref{coalescence} is  represented  
 as,  
 \begin{equation}
  A_i+A_j \rightarrow \left\{
   \begin{array}{ccl}
    A_i & \mbox{if}~~ i-j\equiv 0,1~(\mbox{ mod}~3), \\
    A_j& \mbox{if}~~ i-j\equiv 2~(\mbox{ mod}~3). 
   \end{array} \right. \label{coal3}
   \end{equation}
   There are several  solvable cases for the 
 Smoluchowski coalescence equations \cite{al} for the size of clusters. 
   The "Boltzmann  equation"  for the collision model of 
the extended $2s+1$ types rock-paper-scissors particles is a nonlinear integrable 
 system \cite{bo,  i87, i09, na}.  Consider the time evolution of the relative abundance (concentration ratio), as in \cite{krb}, for 
the "Smoluchowski equation" of the coalescence of extended $2s+1$ types rock-paper-scissors particles.  It is obtained by a logarithmic time change of   the nonlinear integrable system, as will be shown in Section \ref{smo}.

\section{Smoluchowski equation for coalescence of   rock-paper-scissors  particles}\label{rps}

 The Smoluchowski coagulation model is 
for the time evolution  of distribution of cluster  size as Eq. (\ref{cluster}), while our coalescence model given by the following  1), 2), 3) and 4) is for the  distribution of particle types of rock,  paper and scissors. 

\noindent 
Coalescence model of rock-paper-scissors particles:
 
1) There are $3$ types (rock,  paper,  scissors) of particles 
$1,2,3$ whose numbers of particles
at time $t$, are $n_{1}(t), n_{2}(t),n_{3}(t)$ respectively, for which 
$n_{1}(t) + n_{2}(t)  + n_{3}(t) =n(t).$

2) Each of $n(t)$ particles  coalescence  with other particles **$n(t)~ dt$** times on the 
average in $[t,t+dt]$.

3) Each particle is in a chaotic bath of particles.  Each coalescence   
pair is equally likely to be chosen.

4) By a coalescence of  a particle of type $i$
 and a particle of type $j$,  the two particles become one particle of type $i$, if
$ i-j\equiv 0,1~~ (mod~~ 3)$, otherwise become one particle of 
type $j$ as Eq.(\ref{coal3}),
as also shown in Fig. \ref{coalescence}.
\vspace{0.5cm}
 
\begin{figure}[h]
\begin{picture}(420,80)(0,0)

\put(80,75){\line(1,-1){70}}
\put(80,75){\line(-1,-1){70}}

\put(40,35){\line(1,-1){30}}
\put(120,35){\line(-1,-1){30}}

\end{picture}

2\hspace{1.8cm}
3\hspace{0.8cm}
1\hspace{1.5cm}
2\hspace{2.5cm}
\caption{Possible tree of  coalescence model: starting from $n(0)=4$ 
particle. By the coalescence of a particle of  2 (paper) and 
the particle of  3 (scissors) the particle of  3  survives. 
By the coalescence of a particle of  1  (rock) and 
the particle of  2 the particle of  2  survives. 
By  the coalescence of the two survivors
the particle of type 3 (scissors) survives  finally.
}\label{tree1}
\end{figure}
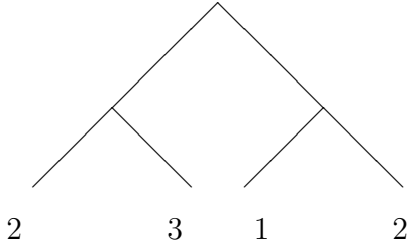

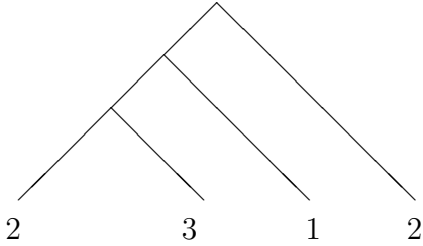
\begin{figure}[h]
\begin{picture}(420,80)(0,0)

\put(80,75){\line(1,-1){75}}
\put(80,75){\line(-1,-1){75}}

\put(60,55){\line(1,-1){55}}
\put(40,35){\line(1,-1){35}}
\end{picture}

2\hspace{2.0cm}
3\hspace{1.3cm}
1\hspace{1.0cm}
2  \\
\caption{Possible tree of  coalescence model:  starting from $n(0)=4$
particle,  by the coalescence of  a particle of  2 (paper)  and 
a particle of 3 (scissors),  the particle of  3  survives, then by the coalescence with the particle of  1 (rock), the particle of  1 survives.  By the coalescence of 
 the  particle of  1 and 
the particle of 2 the particle of  2 (paper)  survives finally.
}\label{tree2}
\end{figure}

Examples of  trees generated by our coalescence model are shown in Fig. \ref{tree1} 
and Fig. \ref{tree2}.
\vspace{.2cm}

Now we derive the master equation.
One of 
particles of type  $1$ and  the other  particle of type $1$ are chosen at random  with probability  $(\frac{n_{1}(t)}{n}dt)n_{1}(t)$ and 
 coalescent with each other  and then become one particle of type 1.
One  particle of type 1 and  one particle of type 3 are chosen at random 
$(\frac{n_{3}(t)}{n}dt)n_{1}(t)$
and 
coalescent  with each other and then become a particle of type 1. 
Consider  
\begin{eqnarray}
  dn_{1}(t)=n_{1}(t+dt)-n_{1}(t). 
\end{eqnarray}
We have 
\begin{eqnarray}
n_{1}(t+dt)=(\frac{1}{2}\frac{n_{1}(t)}{n}n(t)~dt)n_{1}(t)+\frac{n_{3}(t)}{n}n(t)~dt)n_{1}(t)+(1-n(t)~dt) n_1(t).
\end{eqnarray}
Hence we have
\begin{eqnarray}
dn_{1}(t)&=&(\frac{1}{2}\frac{n_{1}(t)}{n}n(t)~dt)n_{1}(t)+\frac{n_{3}(t)}{n}n(t)~dt)n_{1}(t)-n(t)~dt n_1(t)\\
&=& -\frac{1}{2}(n_{1}(t)n_{1}(t)+2n_{1}(t)n_{2}(t))dt,
\end{eqnarray}
which shows  the  master equation
 \begin{equation}\label{Seq}
  \left\{
   \begin{array}{ccl}
   \frac{\partial n_1(t)}{\partial t}&&=-\frac{1}{2}n_1(t) (2 n_2(t) + n_1(t))\\
\frac{\partial n_2(t)}{\partial t}&&= -\frac{1}{2}n_2(t) (2  n_3(t) + n_2(t))\\
\frac{\partial n_3(t)}{\partial t}&&= -\frac{1}{2}n_3(t) (2  n_1(t) + n_3(t)), 
   \end{array} \right. 
   \end{equation}
which is written for Eq. (\ref{coal3}) as, 
\begin{eqnarray}
 \frac{dn_{k}(t)}{dt}= 1/2\sum _{A_{i}+A_{j} \rightarrow A_k } 
  n_{i}n_{j}-\sum_{j=1}^{3}  n_j n_k, 
\end{eqnarray}
which will be studied  for  a general  $2s+1$ hands rock-paper-scissors particles 
in Section \ref{smo}.

Since the coalescence, given in Fig. \ref{coalescence}, is a binary interaction,  from the above   2)  it occurs  $\frac{n(t)}{2}n(t)~ dt$ times in $[t,t+dt]$, as will be mentioned in later Section 
\ref{simulation}.

\section{Collision model of $2s+1$types rock-paper-scissors particles}\label{bol}
Consider the model defined by the following 1), 2), 3), and 4).

\noindent
Collision model of $2s+1$ types rock-paper-scissors particles:

1) There are $2s+1$ types rock-paper-scissors particles 
$1,2,..., 2s+1$ whose abundances
at time $t$, are $n_{1}(t), n_{2}(t),...,n_{2s+1}(t)$ respectively, for which 
$n_{1}(t) + n_{2}(t) +... + n_{2s+1}(t) =n(t).$

2) Each particle collides with other particles $ dt$ times on the 
average per time length $dt$. 

3) Each particle is in a chaotic bath of particles.  Each colliding    
pair is equally likely to be chosen.

4) By a collision  a particle of type $i$ 
 and a particle of type $j$ become two particles of type $i$, if
$ i-j\equiv 0,1,...,s (mod~~ 2s+1)$, otherwise become two particles of 
type $j$ as Eq.(\ref{col}),
as also shown in Fig. \ref{collision} for the case $s=1$.

For the collision model of $2s+1$ types of rock-paper-scissors  particles,  we extend the  rule of the rock-paper-scissors particles  by the cyclic dominance rule, 
 \begin{equation}
  A_i+A_j \rightarrow \left\{
   \begin{array}{ccl}
   2 A_i & \mbox{if}~ i-j\equiv 0,...,s~(\mbox{ mod}~2s+1) \\
   2 A_j& \mbox{if}~ i-j\equiv s+1,...,2s~(\mbox{ mod}~2s+1),
   \end{array} \right. \label{col}
   \end{equation}
 as shown in Fig. \ref{collision} for the case 
$s=1$. 
\vspace{0.5cm}

The total number of particles $n(t)$ is time invariant.  The relative abundance for each type of particles 
 \cite{i71,i75} is given by the master equation 
\begin{eqnarray} 
\frac{\partial }{\partial u}P_i(u)
= P_i(u)(\sum_{j=1}^s P_{i-j}(u) -\sum_{j=1}^{s}P_{i+j}(u)).
\label{integrable}
\end{eqnarray}
 Consider $2r+1$ types  out of the $2s+1$ types $A_1,...,A_{2s+1}$. If each of the 
$2r+1$ types  dominates the other $r$ types  and is dominated by 
the other remained $r$ types, then we say the $2r+1$ types  are 
in a
 regular tournament. 
 
 Take $2r+1$ particles at random  from our system of the $2s+1$ types $A_1,...,A_{2s+1}$ of particles with Eq. (\ref{integrable}). 
Let 
$H_{r}(\vec{P(t)})$ for $\vec{P(t)}\equiv (P_1(t),...,P_{2s+1}(t))$, be the probability  that the  $2r+1$ particles whose 
corresponding $2r+1$ types  
are in a regular tournament. Then we have the $s+1$  conserved quantities  
\cite{i87}
\begin{eqnarray}
H_{r}(\vec{P(t)})=H_{r}(\vec{P(0)}),~~\mbox{for}~~r=0,...,s.
\end{eqnarray}

For example, 
for the case $2s+1=5$, we have the conserved quantities  
\begin{eqnarray}
&&H_{0}(\vec{P(t)})=\sum_{i=1}^{5}P_i(t)\\
&&H_{1}(\vec{P(t)})=\sum_{i=1}^{5}P_i(t)P_{i+1}P_{i+3}\\
&& H_{2}(\vec{P(t)})=P_{1}(t)P_{2}(t)P_{3}(t)P_{4}(t)P_{5}(t).
 \end{eqnarray}

Eq. (\ref{integrable}) is a nonlinear integrable system 
\cite{bo,  i87, i09, na}.

\section{Coalescence  model of $2s+1$-types rock-paper-scissors particles}\label{smo}

We consider the  coalescence model  defined by the following 1), 2), 3) and 4).

\noindent
Coalescence  model of $2s+1$-types rock-paper-scissors particles:

1) There are $2s+1$ types rock-paper-scissors particles 
$1,2,..., 2s+1$ whose abundances
at time $t$, are $n_{1}(t), n_{2}(t),...,n_{2s+1}(t)$ respectively, for which 
$n_{1}(t) + n_{2}(t) +... + n_{2s+1}(t) =n(t).$

2) Each particle coalescence  with other particles $n(t)~ dt$ times on the 
average in  $[t,t+dt]$. 

3) Each particle is in a chaotic bath of particles.  Each coalescence     
pair is equally likely to be chosen.

4) By a coalescence a particle of type $i$ 
 and a particle of type $j$ become one particles of type $i$, if
$ i-j\equiv 0,1,...,s (mod~~ 2s+1)$, otherwise become one particle of 
type $j$ as 
\begin{equation}
  A_i+A_j \rightarrow \left\{
   \begin{array}{ccl}
    A_i & \mbox{if}~ i-j\equiv 0,...,s~(\mbox{ mod}~2s+1) \\
    A_j& \mbox{if}~ i-j\equiv s+1,...,2s~(\mbox{ mod}~2s+1),
   \end{array} \right. \label{coal}
   \end{equation}
   
\vspace{0.3cm}

We have for $i=1,2,...,2s+1$, 
\begin{eqnarray} 
\frac{\partial n_i(t)}{\partial t}=-\frac{1}{2}n_i(t) (n_i(t) +2\sum_{j=1}^{s}n_{i+j}(t))\label{2s+1},
\end{eqnarray}
which shows
\begin{eqnarray} 
\frac{\partial n_i(t)}{\partial t}=\frac{1}{2}n_i(t) (\sum_{j=1}^{s}n_{i-j}(t) -\sum_{j=1}^{s}n_{i+j}(t))-\frac{1}{2}n_i(t)n(t) \label{2s+1'}
\end{eqnarray}
where
$n(t)=n_{1}(t) + n_{2}(t) + \cdots + n_{2s+1}(t) $.

Since 
\begin{eqnarray} 
\frac{\partial n(t)}{\partial t}&&=-\frac{1}{2}n(t)^2,  
\label{riccati} 
\end{eqnarray}
we have
\begin{eqnarray} 
n(t)=\frac{2}{t+2/n(0)}.   
\label{n(t)} 
\end{eqnarray}
As in \cite{krb}, "one useful trick that often simplifies master equations is to eliminate the loss terms by
considering concentration ratios, rather than the concentrations themselves".
We have from Eq. (\ref{2s+1'})
\begin{eqnarray} 
\frac{\partial}{\partial t}\frac{n_i(t)}{n(t)} 
=\frac{1}{2}n(t)\frac{n_i(t)}{n(t)} (\sum_{j=1}^{s}\frac{n_{i-j}(t)}{n(t)} -\sum_{j=1}^{s}\frac{n_{i+j}(t)}{n(t)}). \label{2s+1''}
\end{eqnarray}
Putting  $\frac{n_i(t)}{n(t)}=Q_i(t)$, $i=1,...,2s+1$, 
 we have 
\begin{eqnarray} 
\frac{\partial}{\partial t}Q_i(t)
=\frac{1}{t+\frac{2}{n(0)}} Q_i(t)(\sum_{j=1}^{s}Q_{i-j}(t)
 -\sum_{j=1}^{s}Q_{i+j}(t)). 
\label{2s+1'''}
\end{eqnarray}
For $u=\log [t+\frac{2}{n(0)}]$,  considering
\begin{eqnarray} 
e^u=t+  \frac{2}{n(0)}
\label{eu}
\end{eqnarray}
we have
\begin{eqnarray} 
&&\frac{\partial}{\partial t}Q_i(e^u-  \frac{2}{n(0)})\\
&&=\frac{1}{e^u} Q_i(e^u-  \frac{2}{n(0)})
(\sum_{j=1}^{s}Q_{i-j}(e^u-  \frac{2}{n(0)})\\&& -\sum_{j=1}^{s}Q_{i+j}(e^u-  \frac{2}{n(0)})). 
\label{2s+1''''}
\end{eqnarray}
Since
\begin{eqnarray} 
\frac{\partial u}{\partial t} =\frac{1}{t+\frac{2}{n(0)}}=\frac{1}{e^u}~~ \mbox{and}
~~\frac{\partial}{\partial t}
=\frac{\partial u}{\partial t}\frac{\partial }{\partial u}=\frac{1}{e^u}\frac{\partial }{\partial u},  
\end{eqnarray}
we have 
\begin{eqnarray} 
P_i(u)= Q_i(e^u-  \frac{2}{n(0)}),~~ \mbox{for}~~i=1,...,2s+1. 
\end{eqnarray}
Hence   for the solution $P_i(u)$ to the nonlinear integrable system  Eq. (\ref{integrable}),  we have
\begin{eqnarray} 
Q_i( t)=P_i(\log [t+ \frac{2}{n(0)}]) ,~~ \mbox{for}~~i=1,...,2s+1.\label{log}
\end{eqnarray}
Thus we see the solution of the dynamical system  of the   $2s+1$ types rock-paper-scissors coalescence model is obtained  
by a logarithmic time change of  the solution to  Eq. (\ref{integrable}) for  the  $2s+1$ types rock-paper-scissors collision model.

We can extend our argument to the  infinitely many types rock-paper-scissors particles coalescence model.   Eq. (\ref{integrable})  for the collision model  is extended to 
\begin{eqnarray}
\frac{d}{dt}P(x,t)=P(x,t)\left( \int_{x-\pi}^{x}P(y,t)dy-
\int_{x}^{x+\pi}P(y,t)dy \right)\label{infinite}
\end{eqnarray}
where $P(x,t)=P(x+2\pi,t)$ for each $x,P(x,t)$ being the 
probability density for a point on the unit circle \cite{ev, i88}. 
 The logarithmic time change to Eq. (\ref{infinite}) gives the dynamics of infinitely many  types rock-paper-scissors particles coalescence model.  


\section{Simulation of a discrete time model}\label{simulation}
Consider the coalescence model of rock-paper-scissors 
particles (Section \ref{rps}) with the following 2)' 
instead of 2 in Section \ref{rps}. 

2)'  One of $n(t_l)=n-l$ particles  coalescence  with another particle  
 at the discrete time $t_l$ for $l=0,1,...,n-1$, where 
\begin{eqnarray}  
t_{n-l}= \frac{2}{(n-l)}-\frac{2}{n}~~ \mbox{ for}~~  l=0,1,...,n-1.  \label{timescale}
\end{eqnarray}

 In Section \ref{rps},  from 2) coalescence  occurs  $\frac{n(t)}{2}n(t)~ dt$ times in $(t, t+dt)$ for the total number of particles $n(t)$ at time $t$. Hence for our discrete time model it is natural to  take the above time scale Eq. (\ref{timescale}).

\begin{figure}[h]
\special{epsfile=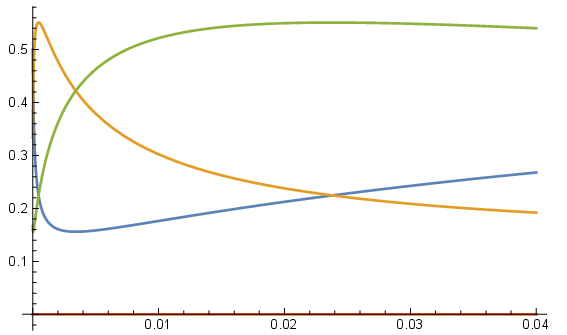 hscale=1.0  vscale=1.0} 
\vspace{6.0cm}
\caption{The numerical solution of Eq. (\ref{2s+1'''})  for $s=1$ starting from the system  with 30000 rock particles (blue), 20000 paper particles(mustard yellow) and 10000 scissors particles (green).}  \label{discrete3}
\end{figure}

\begin{figure}[h]
\special{epsfile=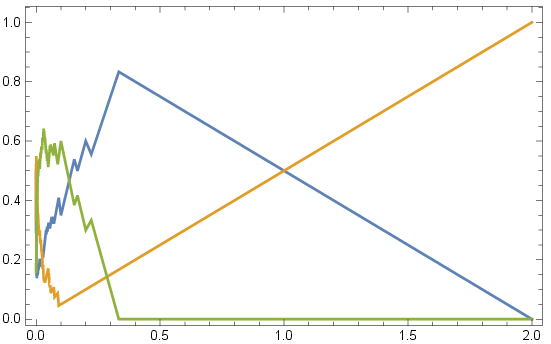 hscale=1.0  vscale=1.0} 
\vspace{6.0cm}
\caption{Starting from the system  with 30000 rock particles, 20000 paper particles and 10000 scissors particles, a  paper particle finally 
survives. }\label{discrete5}
\end{figure}

Since the solution $(n_1(t),n_2(t), n_3(t))$ of  Eq. (\ref{Seq})   approaches to 0 very quickly, here we  show the  relative abundances $(\frac{n_1(t)}{n(t)},\frac{n_2(t)}{n(t)},\frac{n_3(t)}{n(t)})$ in Fig. \ref{discrete3}. 
Eq. (\ref{timescale}) gives the reasonable time scale to compare the simulation results with the solution of the "Smoluchowski equation" Eq. ( \ref{Seq}). 
Starting from the system  with 30000 rock particles, 20000 paper particles and 10000 scissors particles,  a  paper particle finally 
survives as  shown in  Fig. \ref{discrete5}. 

We write $N_i(t)$ for the abundance $i=1,2,3$ for our discrete model where $N_1(t)+N_2(t)+N_3(t)=n(t)$ at time $t$.
As we see in Fig. \ref{discrete3} and Fig. \ref{discrete4}, 
for the period $[0, t_{60000-50}]$, where 50 particles exist at 
$t_{60000-50}=\frac{2}{50}-\frac{2}{60000}$, 
 the trajectory $(\frac{N_1(t)}{n(t)},\frac{N_2(t)}{n(t)},\frac{N_3(t)}{n(t)})$  seems to be  
approximated by the numerical solution of Eq. (\ref{2s+1'''}) qualitatively. As the decrease  of the total number of particles, the fluctuation of the relative abundance of each type increases as we see in Fig. \ref{discrete5}. It is remarkable that at $t_{60000-8}=\frac{2}{8}-\frac{2}{60000}$ all of the three types coexist.  Finally one  paper particle 
survives  at time  $t_{60000-1}=\frac{2}{1}-\frac{2}{60000}$ as shown in Fig. \ref{discrete5}.

For the  system for the collision model (Fig. \ref{collision}) without deterministic approximation, we have a stochastic version of each conserved quantity, which is shown to be  a martingale in probability theory.  Considering the second largest eigen value and its corresponding eigen vector, we obtain the asymptotic probability of 
coexistence \cite{i79} by using the martingale. For our coalescence model 
 (Fig. \ref{coalescence}) to obtain the asymptotic 
probability of coexistence is our next problem. Our simulation result 
(Fig. \ref{discrete5}) may be  an extreme case. However  the three types coexist until very 
final stage in our many other simulations. The conserved quantity  for the deterministic approximation makes the coexistence of rock particles, paper 
particles  and scissors particles for the system with sufficiently large  number of particles. 

\section{Conclusion}\label{conclusion}
\begin{figure}[h]
\special{epsfile=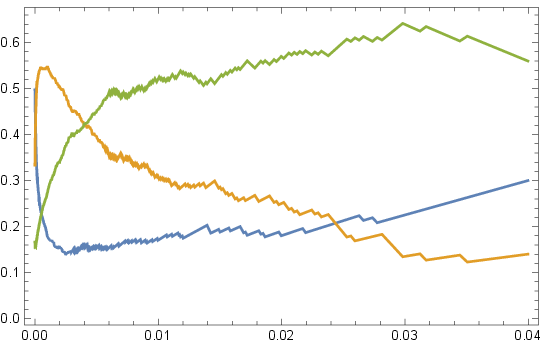 hscale=1.0  vscale=1.0} 
\vspace{6.0cm}
\caption{ The trajectory $(\frac{N_1(t)}{n(t)},\frac{N_2(t)}{n(t)},\frac{N_3(t)}{n(t)})$ for the coalescence model of rock-paper-scissors particles seems to be  
approximated by the numerical solution of Eq. (\ref{2s+1'''})  for $s=1$.}\label{discrete4}
\end{figure}

The "Boltzmann equation" of the collision model of $2s+1$-types rock-paper-scissors particles is shown to be a nonlinear integrable system 
in the previous papers. 
  The relative abundance (concentration ratio) for the solution of 
the "Smoluchowski equation" of the coalescence of  $2s+1$ types rock-paper-scissors particles is obtained by a logarithmic time change of   t the solution of the "Boltzmann equation". 
For the simulation of the rock-paper-scissors coalescence model ( Fig. \ref{discrete4}), the system with sufficiently large number of particles, the 
conserved quantities of our  "Smoluchowski equation"    works for the coexistence of 
three types.
The system with finite number of particles, without deterministic approximation, 
gives interesting probabilistic questions, " which type of particle finally survives?",  "when 
the coexistence of three types of particles ends?", etc. , for our next study.

\clearpage

\end{document}